\documentclass[twocolumn,showpacs,amsmath,amssymb,prl]{revtex4}

\usepackage[dvips]{graphicx}

\usepackage{bm}

\begin{document}
\title{Rejection-free Geometric Cluster Algorithm for Complex Fluids}
\author{Jiwen Liu}
\author{Erik Luijten}
\email[Corresponding author. E-mail: ]{luijten@uiuc.edu}
\affiliation{Department of Materials Science and Engineering, University of
Illinois at Urbana-Champaign, Urbana, Illinois 61801}

\date{September 23, 2003}

\begin{abstract}
  We present a novel, generally applicable Monte Carlo algorithm for the
  simulation of fluid systems.  Geometric transformations are used to identify
  clusters of particles in such a manner that every cluster move is accepted,
  irrespective of the nature of the pair interactions. The rejection-free and
  non-local nature of the algorithm make it particularly suitable for the
  efficient simulation of complex fluids with components of widely varying
  size, such as colloidal mixtures. Compared to conventional simulation
  algorithms, typical efficiency improvements amount to several orders of
  magnitude.
\end{abstract}


\maketitle

The Monte Carlo (MC) method constitutes an important simulation technique in
many areas of physics and chemistry~\cite{binder92,allentildesley87}. One of
its central aspects is the possibility to introduce non-physical dynamics,
permitting the study of systems that evolve over otherwise prohibitively large
time scales. A well-known example is the cluster algorithm for lattice spin
models introduced by Swendsen and Wang (SW)~\cite{swendsen87}, which suppresses
dynamic slowing down near a critical point. Since the conception of this
method, its generalization to off-lattice fluids of interacting particles has
been an elusive goal, the main bottleneck being the absence of particle--hole
symmetry. Also away from the critical point the existence of several different
time and length scales constitutes a major obstacle in the simulation of
complex fluids.  This situation commonly arises in multi-component systems,
such as binary mixtures, colloidal suspensions and colloid--polymer mixtures,
and has essentially precluded the computational study of many such systems.  In
this Letter, we present a novel, rejection-free cluster Monte Carlo method of
considerable generality that alleviates this problem.  It greatly facilitates
the canonical simulation of large classes of continuum systems, such as complex
fluids, by generating particle configurations according to the Boltzmann
distribution, without suffering from severe slowing down in the presence of
large size differences.

The SW lattice cluster algorithm and its improvement by Wolff~\cite{wolff89}
are based upon the Fortuin--Kasteleyn~\cite{fortuin72} mapping of the Potts
model onto the random-cluster model, which decomposes a system of spins (or
Potts variables) into \emph{independent} clusters.
This is manifestly different from collective update schemes in which more or
less arbitrary groups of spins (particles) are flipped (moved).  While such
multiple-particle moves have yielded significant improvements in specific
situations~\cite{wu92,jaster99,woodcock99,lobaskin99}, their acceptance rate
generally is an exponentially decreasing function of the number of particles
involved. By contrast, the SW algorithm is rejection-free: every completed
cluster is flipped without an additional evaluation of the resulting energy
difference.  Efficient \emph{off-lattice} cluster algorithms are only known for
the Widom--Rowlinson and Stillinger--Helfand models for fluid
mixtures~\cite{johnson97,sun00}, in which identical particles do not interact
at all.  Furthermore, for hard-sphere fluids Dress and Krauth~\cite{dress95}
have proposed a cluster algorithm based upon geometric operations, that is
capable of relaxing size-asymmetric mixtures~\cite{buhot98} and model glass
formers~\cite{santen00}. However, hitherto no general off-lattice equivalent of
the SW approach has been developed.  In this paper, we demonstrate that the
geometric method can be formulated as a Wolff single-cluster algorithm and
subsequently be extended to arbitrary pair potentials between the constituents,
while maintaining its rejection-free nature. The resulting \emph{generalized}
geometric cluster algorithm (GCA) handles interactions in the same manner as
the SW algorithm, and in this respect can be considered as its counterpart for
continuum systems.  Lattice cluster methods as well as the GCA exploit
invariance of the Hamiltonian under global symmetry operations.

\begin{figure*}[t]
\begin{center}
\includegraphics[width=0.8\textwidth]{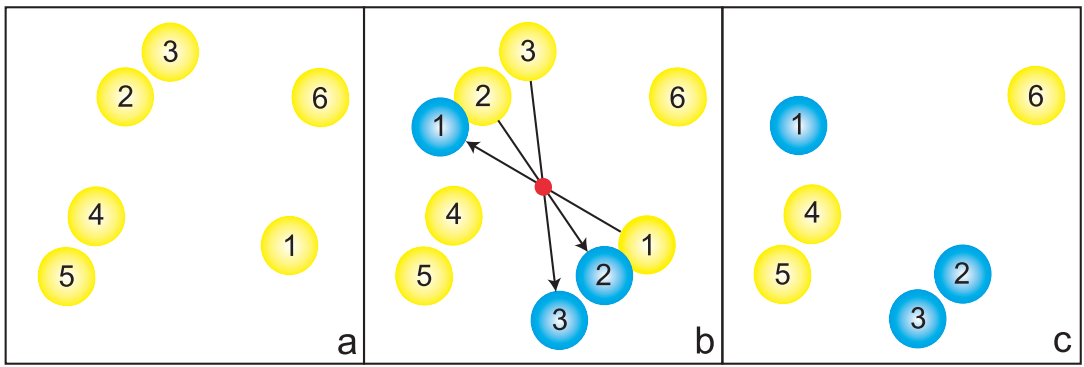}
\caption{Two-dimensional illustration of the interacting geometric
cluster algorithm. Light and dark colors label the particles before and after
the geometrical operation, respectively. The small disk denotes the pivot.
a)~Initial configuration; b)~construction of a new cluster via point reflection
of particles $1$--$3$ with respect to the pivot; c)~final configuration.}
\label{fig:geomc}
\end{center}
\end{figure*}

In the original GCA~\cite{dress95} a molecular configuration is rotated around
an arbitrarily chosen pivot and overlaid with its original (non-rotated)
version.  Objects that overlap between the original and rotated configurations
lead to ``clusters'' of particles. The particles belonging to these clusters
are exchanged between the original and the rotated configuration.  Since the
pair potential is either zero or infinity, each configuration without particle
overlaps has the same Boltzmann factor and hence the same probability. It has
been suggested~\cite{dress95,malherbe99} to extend this approach to other pair
potentials by introducing a Metropolis-type criterion for the acceptance of a
cluster move. However, this approach faces serious drawbacks.  (i)~It cannot be
applied to soft-core potentials, since the cluster-construction process fails
to generate configurations containing interpenetrating potentials. (ii)~The
efficiency strongly deteriorates, 
since the algorithm is no longer rejection-free. Indeed, the stronger the
interactions, the less relevant the (athermal) clusters become, and for many
practical cases a conventional single-particle Metropolis algorithm will be
more efficient~\cite{pastore02}.  It is noteworthy that for lattice spin models
the original GCA has successfully been generalized to include attractive
short-range interactions~\cite{heringa98}.  However, this approach implicitly
exploits particle--hole symmetry by relying on a mapping between sites in the
original and the rotated lattice structure.

In order to formulate a geometric cluster algorithm for interacting fluids, we
first rephrase the cluster construction for the original GCA as follows. After
a random pivot has been chosen, the first particle is picked at random and
moved via a point reflection with respect to the pivot (this reflection
replaces the rotation). If this leads to one or more overlaps, the
corresponding particles are also moved with respect to the same pivot.  This
procedure is reiterated recursively until no more overlaps are present. For the
next cluster, a new pivot is chosen.  In the presence of a general pair
potential $V(r)$, we generalize this scheme as illustrated in
Fig.~\ref{fig:geomc}.  After the first particle~$i$ has been moved from
position~$\mathbf{r}_i$ to its new position~$\mathbf{r}_i'$, two classes of
particles are identified: (a) particles that interact with $i$ in its
\emph{original} position; (b) particles that interact with $i$ in its
\emph{new} position. Every particle that belongs to category (a) or~(b) is
subsequently considered for inclusion in the cluster, i.e., for reflection with
respect to the pivot. Particles that fall into both categories are considered
only once.  While the first particle~$i$ is always moved, subsequent
particles~$j$ are added to the cluster with a probability $p_{ij} = \max[1 -
\exp(-\beta \Delta_{ij}), 0]$, where $\Delta_{ij} =
V(|\mathbf{r}_i'-\mathbf{r}_j|) - V(|\mathbf{r}_i-\mathbf{r}_j|)$ and $\beta =
1/k_{\rm B}T$. Thus, the cluster addition probability for particle~$j$
\emph{solely} depends on the energy difference corresponding to a change in
relative position of $i$ and~$j$. Other energy differences resulting from a
move of particle~$j$ are \emph{not} taken into account, which distinguishes
this method from a standard Metropolis algorithm with multiple particle moves
and makes it the analog of the Wolff cluster algorithm.  Instead, the procedure
is carried out iteratively. If particle~$j$ is added to the cluster, then all
its interacting neighbors (both in category~(a) and in category~(b)) that have
not yet been added to the cluster are considered for inclusion as well. The
cluster construction is completed once all interacting neighbors have been
considered.

A particle~$j$ in category~(a) that is added to the cluster can be viewed as
``moving with'' particle~$i$; a particle~$j$ in category~(b) is then
interpreted as ``moving from'' particle~$i$ in its new position. However, in
either case particles $i$ and~$j$ maintain their original separation. The
system evolves by virtue of the particles that are \emph{not} included in the
cluster. Again, this is analogous to a spin cluster algorithm, in which all
spins within a given cluster maintain their relative orientation. In the limit
of a pure hard-core repulsion, category~(a) particles do not exist and the
addition probability for category~(b) particles is unity. Thus, the original
GCA~\cite{dress95} indeed constitutes a special case of the generalized GCA.

The ergodicity of this algorithm follows from the fact that there is a
non-vanishing probability that a cluster consists of only one particle, which
can be moved over an arbitrarily small distance, since the location of the
pivot is chosen at random.  Despite the presence of a variable addition
probability and the existence of two categories of particle moves, the proof of
detailed balance proceeds in a similar way as for the Wolff
algorithm~\cite{wolff89}.  In the transition from a given configuration~$X$
(energy $E_X$) to a new configuration~$Y$ (energy $E_Y$), an energy change is
induced by every interacting particle that is \emph{not} added to the cluster.
The probability of such a ``broken bond''~$k$ is given by $1-p_{k}$, which is
unity if the \emph{pair energy} decreases. This is the analog of a pair of
\emph{antiparallel} spins in a lattice cluster algorithm.  For an
\emph{increase} in pair energy~$\Delta_{k}$, the probability of breaking a bond
is $\exp(-\beta \Delta_{k})$. Accordingly, the creation of a certain cluster
corresponds to breaking a number of bonds, which has a probability
\begin{equation}
T(X\to Y) =
 C \prod_{k} (1-p_k) = C \exp\left[-\beta \sum_{l} \Delta_l\right] \;,
\end{equation}
where the product runs over the set~$\{k\}$ of all \emph{broken} bonds, which
is comprised of the subset $\{l\}$ of broken bonds~$l$ that lead to an increase
in pair energy and the subset $\{m\}$ of broken bonds that lead to a decrease
in pair energy.  The factor $C$ accounts for the probability of creating a
specific arrangement of bonds inside the cluster. The probability for the
reverse move runs over the same set $\{k\}$, but all energy differences have
changed sign (indicated by $\bar{p}$) and the sum over $\Delta_l$ is replaced
by the negative sum over the complementary set $\{m\}$,
\begin{equation}
T(Y\to X) =
 C \prod_{k} (1-\bar{p}_k) = C \exp\left[+ \beta \sum_{m} \Delta_m\right] \;.
\end{equation}
The probability of picking a specific particle as the starting point for this
cluster is identical in the forward and the reverse move. Moreover, we require
the geometric operation to be self-inverse.  For clusters thus constructed, we
then indeed have succeeded in fulfilling detailed balance while maintaining an
acceptance ratio of unity:
\begin{equation}
\frac{T(X\to Y)}{T(Y\to X)} =
 \exp\left[ -\beta \sum_{k} \Delta_k\right] =
 \frac{\exp(-\beta E_{Y})}{\exp(-\beta E_{X})}
 \;.
\end{equation}

\begin{figure}
\includegraphics[width=0.4\textwidth]{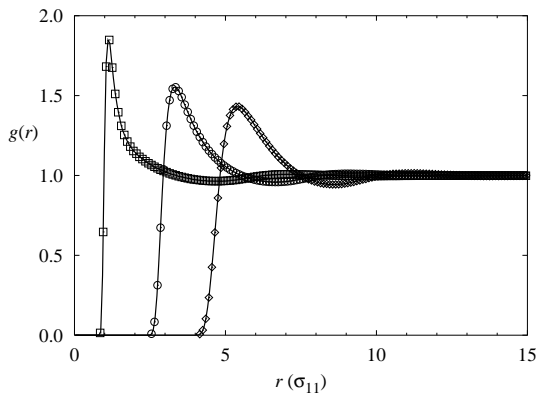}
\caption{Comparison between a conventional molecular dynamics calculation
(solid lines) and the geometric cluster algorithm (symbols), for a
size-asymmetric binary Lennard-Jones mixture. Shown are, from left to right,
the correlation functions for small--small, large--small, and large--large
pairs. The agreement is clearly excellent.}
\label{fig:md}
\end{figure}

Figure~\ref{fig:md} illustrates the agreement between the generalized GCA and a
conventional NVT molecular dynamics simulation for a binary Lennard-Jones
mixture containing $800$ small and $400$ large particles at a total packing
fraction $\eta \approx 0.213$. The respective particle diameters are
$\sigma_{11} = 1.0$ and $\sigma_{22} = 5.0$ and the interaction strengths equal
$\varepsilon_{11} = 0.40$ and $\varepsilon_{22} = 0.225$, supplemented by the
Lorentz--Berthelot mixing rules~\cite{allentildesley87}. The particles are
contained in a cubic cell with periodic boundary conditions. All interactions
are cut off at $3\sigma_{22}$.
Evidently, the GCA is capable of handling soft-core potentials.

\begin{figure}
\includegraphics[width=0.4\textwidth]{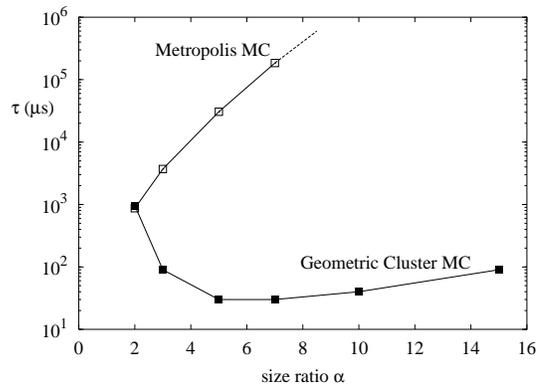}
\caption{Efficiency comparison between a conventional local update algorithm
(open symbols) and the generalized geometric cluster algorithm (closed
symbols), for a size-asymmetric binary mixture of Yukawa particles. As opposed
to the local algorithm, the autocorrelation time per particle (expressed in
$\mu$s of CPU time) for the GCA depends only weakly on size ratio $\alpha$
(variations correspond to changes in the volume ratio of large \emph{vs} small
particles in the cluster), resulting in an efficiency improvement of several
orders of magnitude already for moderate~$\alpha$.}
\label{fig:efficiency}
\end{figure}

The true advantage of the generalized GCA transpires upon consideration of its
efficiency. As a simple model system with intrinsically slow dynamics we again
consider a binary fluid mixture of $N_1$ small and $N_2$ large spherical
particles with size ratio $\alpha \equiv \sigma_{22} / \sigma_{11} \ge 1$. The
particles are contained in a fixed volume, at equal packing fractions $\eta_1 =
\eta_2 = 0.1$. $N_2$ is fixed at $150$ and $N_1$ increases from $1\,200$ to
$506\,250$ as $\alpha$ is varied from $2$ to~$15$.  While pairs of small
particles, as well as pairs involving a large and a small particle, act like
hard spheres, the large particles have a Yukawa repulsion,
\begin{equation}
\label{eq:yu}
U_{22}(r) = \left \{ \begin{array}{ll}
  +\infty                                            & r \le \sigma_{22}\\
  J \exp[-\kappa (r-\sigma_{22})]/(r/ \sigma_{22})   & r > \sigma_{22}  \;,
  \end{array} \right.
\end{equation}
where $\beta J = 3.0$ and the screening length $\kappa^{-1} = \sigma_{11}$.
The Hamiltonian describing the system is given by the sum over all pair
interactions.
As a measure of efficiency we consider the integrated autocorrelation
time~$\tau$ for the energy~\cite{binder01}.
For conventional MC calculations, $\tau$ rapidly increases with
increasing~$\alpha$, because large particles tend to get trapped by particles
belonging to the smaller species (this situation will further deteriorate in
the presence of an attraction between large and small particles). Indeed, for
$\alpha > 7$ it was not even feasible to accurately estimate $\tau$ within a
reasonable amount of CPU time.  By contrast, the generalized GCA has an
autocorrelation time that only weakly depends on the size ratio, as illustrated
in Fig.~\ref{fig:efficiency}. At $\alpha=7$ the resulting efficiency gain
already amounts to more than three orders of magnitude.

\begin{figure}
\includegraphics[width=0.4\textwidth]{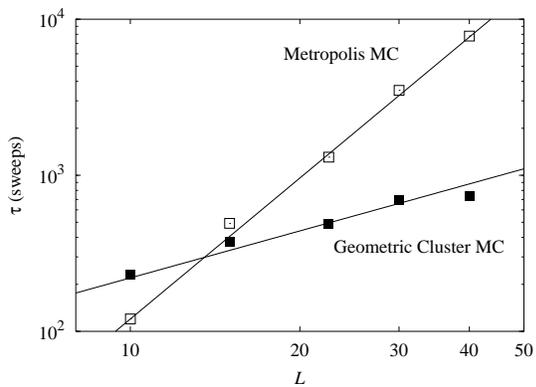}
\caption{Energy autocorrelation times $\tau$ \emph{vs} linear
system size for a critical Lennard-Jones fluid, in units of particle sweeps.}
\label{fig:lj}
\end{figure}

To explore the performance of our algorithm near a critical point, we have
simulated the one-component Lennard-Jones fluid with a potential cutoff $r_{\rm
c}=2.5\sigma$ at $T^{*} = k_{B}T/\varepsilon = 1.19$ and $\rho^* = \rho\sigma^3
= 0.3197$, very close to criticality~\cite{wilding95}.  The energy
autocorrelation time $\tau$ exhibits a power-law dependence on the linear
system size~$L$ for local moves as well as cluster moves, see
Fig.~\ref{fig:lj}.  Since the density is a conserved quantity,
\emph{hydrodynamic} slowing down must be anticipated~\cite{hohenberg77}. Owing
to field mixing~\cite{bruce92}, this will also manifest itself in the critical
energy correlations. Thus, the acceleration $\sim L^{2.1}$ achieved by the GCA
cannot unequivocally be ascribed to the (partial) suppression of critical
slowing
down. 
This is consistent with the observation that the average relative cluster size
grows faster than $L^{\gamma/\nu}$. Given the number of factors that determine
the cluster-growth process, care must be taken to generalize these observations
to the performance of the GCA at the critical point of other fluids.

Suppression of critical slowing down is the primary benefit of cluster
algorithms for lattice systems, making it a crucial requirement that the
percolation threshold of the cluster-formation process coincides with the
critical point. The generalized GCA, on the other hand, addresses a much larger
class of problems by accelerating fluid simulations over a wide range of
temperatures and packing fractions. Its essential limitation is that the
clusters must occupy only part of the system. The average cluster size depends
not only on the interaction strength, but also on the total packing fraction
and size and shape of all constituents. Although no unique percolation
threshold can be defined in a continuum system of interacting particles, we
have observed that the average relative cluster size increases abruptly above a
certain packing fraction, rapidly lowering the computational efficiency.  This
packing fraction depends on temperature and system properties, but $\eta
\approx 0.23$--$0.25$ represents a typical threshold. For increasing size ratio
or degree of polydispersity, we expect the window of accessible packing
fractions to grow, in accordance with the increase of the percolation threshold
as a function of polydispersity~\cite{mecke02}. Indeed, for the binary mixtures
of Fig.~\ref{fig:efficiency}, the relative cluster size rapidly decreases with
increasing~$\alpha$ at fixed total packing fraction.

In summary, we have introduced the first general rejection-free cluster
algorithm for off-lattice systems. Its premier significance lies in a
performance increase of many orders of magnitude for complex fluids in which
the constituents exhibit a large size asymmetry, thus enabling the simulation
of mixtures that were hitherto only accessible via an effective one-component
approach. Size ratios up to~$100$ have been reached in simulations involving
several millions of particles. Our approach can be extended in several ways,
including the treatment of non-spherical particles and electrostatic
interactions.

\begin{acknowledgments}
  This material is based upon work supported by the U.S. Department of Energy,
  Division of Materials Sciences under Award No.\ DEFG02-91ER45439, through the
  Frederick Seitz Materials Research Laboratory at the University of Illinois
  at Urbana-Champaign, and by the STC Program of the National Science
  Foundation under Agreement No.\ CTS-0120978.
\end{acknowledgments}


\begin{thebibliography}{22}
\expandafter\ifx\csname natexlab\endcsname\relax\def\natexlab#1{#1}\fi
\expandafter\ifx\csname bibnamefont\endcsname\relax
  \def\bibnamefont#1{#1}\fi
\expandafter\ifx\csname bibfnamefont\endcsname\relax
  \def\bibfnamefont#1{#1}\fi
\expandafter\ifx\csname citenamefont\endcsname\relax
  \def\citenamefont#1{#1}\fi
\expandafter\ifx\csname url\endcsname\relax
  \def\url#1{\texttt{#1}}\fi
\expandafter\ifx\csname urlprefix\endcsname\relax\def\urlprefix{URL }\fi
\providecommand{\bibinfo}[2]{#2}
\providecommand{\eprint}[2][]{\url{#2}}

\bibitem[{\citenamefont{Binder}(1992)}]{binder92}
\bibinfo{editor}{\bibfnamefont{K.}~\bibnamefont{Binder}}, ed.,
  \emph{\bibinfo{title}{The Monte Carlo Method in Condensed Matter Physics}},
  vol.~\bibinfo{volume}{71} of \emph{\bibinfo{series}{Topics in Applied
  Physics}} (\bibinfo{publisher}{Springer}, \bibinfo{address}{Berlin},
  \bibinfo{year}{1992}).

\bibitem[{\citenamefont{Allen and Tildesley}(1987)}]{allentildesley87}
\bibinfo{author}{\bibfnamefont{M.~P.} \bibnamefont{Allen}} \bibnamefont{and}
  \bibinfo{author}{\bibfnamefont{D.~J.} \bibnamefont{Tildesley}},
  \emph{\bibinfo{title}{Computer Simulations of Liquids}}
  (\bibinfo{publisher}{Clarendon}, \bibinfo{address}{Oxford},
  \bibinfo{year}{1987}).

\bibitem[{\citenamefont{Swendsen and Wang}(1987)}]{swendsen87}
\bibinfo{author}{\bibfnamefont{R.~H.} \bibnamefont{Swendsen}} \bibnamefont{and}
  \bibinfo{author}{\bibfnamefont{J.-S.} \bibnamefont{Wang}},
  \bibinfo{journal}{Phys. Rev. Lett.} \textbf{\bibinfo{volume}{58}},
  \bibinfo{pages}{86} (\bibinfo{year}{1987}).

\bibitem[{\citenamefont{Wolff}(1989)}]{wolff89}
\bibinfo{author}{\bibfnamefont{U.}~\bibnamefont{Wolff}},
  \bibinfo{journal}{Phys. Rev. Lett.} \textbf{\bibinfo{volume}{62}},
  \bibinfo{pages}{361} (\bibinfo{year}{1989}).

\bibitem[{\citenamefont{Fortuin and Kasteleyn}(1972)}]{fortuin72}
\bibinfo{author}{\bibfnamefont{C.~M.} \bibnamefont{Fortuin}} \bibnamefont{and}
  \bibinfo{author}{\bibfnamefont{P.~W.} \bibnamefont{Kasteleyn}},
  \bibinfo{journal}{Physica} \textbf{\bibinfo{volume}{57}},
  \bibinfo{pages}{536} (\bibinfo{year}{1972}).

\bibitem[{\citenamefont{Wu et~al.}(1992)\citenamefont{Wu, Chandler, and
  Smit}}]{wu92}
\bibinfo{author}{\bibfnamefont{D.}~\bibnamefont{Wu}},
  \bibinfo{author}{\bibfnamefont{D.}~\bibnamefont{Chandler}}, \bibnamefont{and}
  \bibinfo{author}{\bibfnamefont{B.}~\bibnamefont{Smit}}, \bibinfo{journal}{J.
  Phys. Chem.} \textbf{\bibinfo{volume}{96}}, \bibinfo{pages}{4077}
  (\bibinfo{year}{1992}).

\bibitem[{\citenamefont{Jaster}(1999)}]{jaster99}
\bibinfo{author}{\bibfnamefont{A.}~\bibnamefont{Jaster}},
  \bibinfo{journal}{Physica A} \textbf{\bibinfo{volume}{264}},
  \bibinfo{pages}{134} (\bibinfo{year}{1999}).

\bibitem[{\citenamefont{Lue and Woodcock}(1999)}]{woodcock99}
\bibinfo{author}{\bibfnamefont{L.}~\bibnamefont{Lue}} \bibnamefont{and}
  \bibinfo{author}{\bibfnamefont{L.~V.} \bibnamefont{Woodcock}},
  \bibinfo{journal}{Mol. Phys.} \textbf{\bibinfo{volume}{96}},
  \bibinfo{pages}{1435} (\bibinfo{year}{1999}).

\bibitem[{\citenamefont{Lobaskin and Linse}(1999)}]{lobaskin99}
\bibinfo{author}{\bibfnamefont{V.}~\bibnamefont{Lobaskin}} \bibnamefont{and}
  \bibinfo{author}{\bibfnamefont{P.}~\bibnamefont{Linse}}, \bibinfo{journal}{J.
  Chem. Phys.} \textbf{\bibinfo{volume}{111}}, \bibinfo{pages}{4300}
  (\bibinfo{year}{1999}).

\bibitem[{\citenamefont{Johnson et~al.}(1997)\citenamefont{Johnson, Gould,
  Machta, and Chayes}}]{johnson97}
\bibinfo{author}{\bibfnamefont{G.}~\bibnamefont{Johnson}},
  \bibinfo{author}{\bibfnamefont{H.}~\bibnamefont{Gould}},
  \bibinfo{author}{\bibfnamefont{J.}~\bibnamefont{Machta}}, \bibnamefont{and}
  \bibinfo{author}{\bibfnamefont{L.~K.} \bibnamefont{Chayes}},
  \bibinfo{journal}{Phys. Rev. Lett.} \textbf{\bibinfo{volume}{79}},
  \bibinfo{pages}{2612} (\bibinfo{year}{1997}).

\bibitem[{\citenamefont{Sun et~al.}(2000)\citenamefont{Sun, Gould, Machta, and
  Chayes}}]{sun00}
\bibinfo{author}{\bibfnamefont{R.}~\bibnamefont{Sun}},
  \bibinfo{author}{\bibfnamefont{H.}~\bibnamefont{Gould}},
  \bibinfo{author}{\bibfnamefont{J.}~\bibnamefont{Machta}}, \bibnamefont{and}
  \bibinfo{author}{\bibfnamefont{L.~W.} \bibnamefont{Chayes}},
  \bibinfo{journal}{Phys. Rev. E} \textbf{\bibinfo{volume}{62}},
  \bibinfo{pages}{2226} (\bibinfo{year}{2000}).

\bibitem[{\citenamefont{Dress and Krauth}(1995)}]{dress95}
\bibinfo{author}{\bibfnamefont{C.}~\bibnamefont{Dress}} \bibnamefont{and}
  \bibinfo{author}{\bibfnamefont{W.}~\bibnamefont{Krauth}},
  \bibinfo{journal}{J. Phys. A} \textbf{\bibinfo{volume}{28}},
  \bibinfo{pages}{L597} (\bibinfo{year}{1995}).

\bibitem[{\citenamefont{Buhot and Krauth}(1998)}]{buhot98}
\bibinfo{author}{\bibfnamefont{A.}~\bibnamefont{Buhot}} \bibnamefont{and}
  \bibinfo{author}{\bibfnamefont{W.}~\bibnamefont{Krauth}},
  \bibinfo{journal}{Phys. Rev. Lett.} \textbf{\bibinfo{volume}{80}},
  \bibinfo{pages}{3787} (\bibinfo{year}{1998}).

\bibitem[{\citenamefont{Santen and Krauth}(2000)}]{santen00}
\bibinfo{author}{\bibfnamefont{L.}~\bibnamefont{Santen}} \bibnamefont{and}
  \bibinfo{author}{\bibfnamefont{W.}~\bibnamefont{Krauth}},
  \bibinfo{journal}{Nature} \textbf{\bibinfo{volume}{405}},
  \bibinfo{pages}{550} (\bibinfo{year}{2000}).

\bibitem[{\citenamefont{Malherbe and Amokrane}(1999)}]{malherbe99}
\bibinfo{author}{\bibfnamefont{J.~G.} \bibnamefont{Malherbe}} \bibnamefont{and}
  \bibinfo{author}{\bibfnamefont{S.}~\bibnamefont{Amokrane}},
  \bibinfo{journal}{Mol. Phys.} \textbf{\bibinfo{volume}{97}},
  \bibinfo{pages}{677} (\bibinfo{year}{1999}).

\bibitem[{\citenamefont{Mbamala and Pastore}(2002)}]{pastore02}
\bibinfo{author}{\bibfnamefont{E.~C.} \bibnamefont{Mbamala}} \bibnamefont{and}
  \bibinfo{author}{\bibfnamefont{G.}~\bibnamefont{Pastore}},
  \bibinfo{journal}{Physica A} \textbf{\bibinfo{volume}{313}},
  \bibinfo{pages}{312} (\bibinfo{year}{2002}).

\bibitem[{\citenamefont{Heringa and Bl{\"o}te}(1998)}]{heringa98}
\bibinfo{author}{\bibfnamefont{J.~R.} \bibnamefont{Heringa}} \bibnamefont{and}
  \bibinfo{author}{\bibfnamefont{H.~W.~J.} \bibnamefont{Bl{\"o}te}},
  \bibinfo{journal}{Phys. Rev. E} \textbf{\bibinfo{volume}{57}},
  \bibinfo{pages}{4976} (\bibinfo{year}{1998}).

\bibitem[{\citenamefont{Binder and Luijten}(2001)}]{binder01}
\bibinfo{author}{\bibfnamefont{K.}~\bibnamefont{Binder}} \bibnamefont{and}
  \bibinfo{author}{\bibfnamefont{E.}~\bibnamefont{Luijten}},
  \bibinfo{journal}{Physics Reports} \textbf{\bibinfo{volume}{344}},
  \bibinfo{pages}{179} (\bibinfo{year}{2001}).

\bibitem[{\citenamefont{Wilding}(1995)}]{wilding95}
\bibinfo{author}{\bibfnamefont{N.~B.} \bibnamefont{Wilding}},
  \bibinfo{journal}{Phys. Rev. E} \textbf{\bibinfo{volume}{52}},
  \bibinfo{pages}{602} (\bibinfo{year}{1995}).

\bibitem[{\citenamefont{Hohenberg and Halperin}(1977)}]{hohenberg77}
\bibinfo{author}{\bibfnamefont{P.~C.} \bibnamefont{Hohenberg}}
  \bibnamefont{and} \bibinfo{author}{\bibfnamefont{B.~I.}
  \bibnamefont{Halperin}}, \bibinfo{journal}{Rev. Mod. Phys.}
  \textbf{\bibinfo{volume}{49}}, \bibinfo{pages}{435} (\bibinfo{year}{1977}).

\bibitem[{\citenamefont{Bruce and Wilding}(1992)}]{bruce92}
\bibinfo{author}{\bibfnamefont{A.~D.} \bibnamefont{Bruce}} \bibnamefont{and}
  \bibinfo{author}{\bibfnamefont{N.~B.} \bibnamefont{Wilding}},
  \bibinfo{journal}{Phys. Rev. Lett.} \textbf{\bibinfo{volume}{68}},
  \bibinfo{pages}{193} (\bibinfo{year}{1992}).

\bibitem[{\citenamefont{Mecke and Seyfried}(2002)}]{mecke02}
\bibinfo{author}{\bibfnamefont{K.~R.} \bibnamefont{Mecke}} \bibnamefont{and}
  \bibinfo{author}{\bibfnamefont{A.}~\bibnamefont{Seyfried}},
  \bibinfo{journal}{Europhys. Lett.} \textbf{\bibinfo{volume}{58}},
  \bibinfo{pages}{28} (\bibinfo{year}{2002}).

\end{thebibliography}

\end{document}